\documentclass[english]{article}
\usepackage[latin9]{inputenc}
\usepackage{geometry}
\usepackage{amstext}
\usepackage{graphicx}
\usepackage{setspace}
\usepackage{babel}
\usepackage{graphicx}
\usepackage{times}
\begin{document}

\title{Clonal interference and Muller's ratchet in spatial habitats}

\author{Jakub Otwinowski\footnote{Emory University, Physics Department Atlanta, Georgia, United States} \footnote{University of Pennsylvania, Biology Department, Philadelphia, Pennsylvania, United States}, Joachim Krug\footnote{University of Cologne, Institute for Theoretical Physics, K\"oln, Germany}}

\maketitle

\abstract{Competition between independently arising beneficial mutations is enhanced in spatial populations due to the linear rather than exponential growth of clones. Recent theoretical studies have pointed out that the resulting fitness dynamics is analogous to a surface growth process, where new layers nucleate and spread stochastically, leading to the build up of scale-invariant roughness. This scenario differs qualitatively from the standard view of adaptation in that the speed of adaptation becomes independent of population size while the fitness variance does not. Here we exploit recent progress in the understanding of surface growth processes to obtain precise predictions for the universal, non-Gaussian shape of the fitness distribution for one-dimensional habitats, which are verified by simulations. When the mutations are deleterious rather than beneficial the problem becomes a spatial version of Muller's ratchet. In contrast to the case of well-mixed populations, the rate of fitness decline remains finite even in the limit of an infinite habitat, provided the ratio $U_d/s^2$ between the deleterious mutation rate and the square of the (negative) selection coefficient is sufficiently large. Using again an analogy to surface growth models we show that the transition between the stationary and the moving state of the ratchet is governed by directed percolation.}

\normalsize

\section{Introduction}

The appearance of a mutation in a population and its fixation or loss
is the most basic process of adaptation. This process determines the
rate of evolution, or how quickly populations adapt to new environments.
One approach in quantitative models of adaptation is to assume that non-neutral mutations are rare. In this regime, generally referred to as periodic selection \cite{Atwood1951,deVisser2006}, the population has no genetic variation except for brief periods when a mutation sweeps through and fixates, and therefore the rate of fixation is mutation limited. Adaptation in the regime of periodic selection and strong selection, dubbed strong selection weak mutation (SSWM) has been studied extensively in the context of extreme value theory \cite{Gillespie1994}. Notably, these models assumed populations were well-mixed, with no spatial structure.

When mutations are more common due to higher rates or larger
population sizes, genetic variation builds up. Clonal interference is
the competition between these mutations to reach high frequency, when
there is little or no recombination. This is especially relevant in
the case of beneficial mutations in microbial populations, since recent experiments suggest they
are more common than previously thought \cite{Desai2007a,Perfeito2007,Sniegowski2010}. Microbial evolution experiments have observed reduced rates of evolution due to the competition between beneficial mutations \cite{deVisser1999,Shaver2002}. Fisher's fundamental theorem
equates the rate of evolution with the variance of the fitness distribution \cite{Fisher1930},
which can be approximated analytically in simplified population genetic
models. These recent theoretical analyses have found the rate of evolution
in large populations of asexuals is not proportional to the total supply rate 
of beneficial mutations, but depends much more weakly (logarithmically) on population size and mutation rate
\cite{Desai2007,Rouzine2008,Park2010a,Yu2010,Good2012,Kelly2013}. 

Deleterious mutations are more common than beneficial ones, but their
chance of fixation is much smaller and vanishes for infinite populations. However, 
in finite populations and in the absence of beneficial
mutations and recombination, deleterious mutations will eventually fix
by genetic drift, leading to a fitness decline known as Muller's
ratchet \cite{Muller1964,Haigh1978}. Determining the rate of the
ratchet as a function of population size, mutation rate and selection
strength is a long-standing problem that continues to attract considerable interest \cite{Stephan1993,Gordo2000,Gordo2000a,Rouzine2008,Jain2008,EtheridgeA2009,Waxman2010,Neher2012a,Metzger2013}.
% Recombination is typically absent in these models. 
Recombination can prevent Muller's ratchet and also mitigates the slowdown in the rate of evolution from clonal interference, which is why Muller's ratchet and clonal interference are often argued as reasons for an evolutionary advantage of sex \cite{Muller1932,Charlesworth2012,Park2013}.

Previous analyses of clonal interference and Muller's ratchet 
were largely limited to well-mixed populations, where
each individual competes with the whole population, such as microbes
in liquid culture. However, many populations are not well-mixed, but are confined
in space such that they only compete with a limited neighborhood population
on timescales of a generation. Spatially structured population genetics have been studied with finitely subdivided, and continuous populations \cite{Fisher1937,Kolmogorov1937,Wright1943,Kimura1964,Malecot1975,Korolev2010}. In a spatially structured population individuals compete only within a limited spatial neighborhood, reducing the effective population size. However, when mutations are rare a single beneficial mutation can compete with the whole population, and the fixation probability is the same in well-mixed and spatially structured populations \cite{Maruyama1970,Maruyama1974}.
Recently, models incorporating large amounts of beneficial mutations
and one or two dimensional spatial structure have found the rate of
evolution to be even slower than in well-mixed populations, as the
slower than exponential growth of clones increases the likelihood of
competition
\cite{Gordo2006,Perfeito2006,Otwinowski2011a,Martens2011b,Claudino2013,Pokalyuk2013}. In
fact, the rate of evolution becomes independent of system size, while
the variance (in the steady state) scales as a power of population
size, violating Fisher's theorem
\cite{Otwinowski2011a,Martens2011b}. This also implies that there is a
long transient regime during which the stationary variance builds up,
while the speed of adaptation is constant. 

Here, we study the transient regime of Wright-Fisher models of
evolution on a one-dimensional lattice for both adaptation with
beneficial mutations, and Muller's ratchet. 
In the former case the fitness variance grows as a power law in time,
and saturates at a value determined by the system size (also as a
power law) \cite{Otwinowski2011a,Martens2011b}. The fitness may be
pictured as a surface in space, which advances over time. This
behavior is analogous to surface growth models in physics, where
particles are deposited on an initially flat surface, which develops
roughness over time \cite{Barabasi1995,Krug1997}. Specifically, the
accumulation of beneficial mutations was found to be analogous to
polynuclear growth in ref.~\cite{Martens2011b}. In the next section we introduce the model for adaptation on a one dimensional lattice, and review previous results. In section \ref{adaptation}, we use extensive simulations to show the model of adaptation belongs to a class of surface growth models called the Kardar-Parisi-Zhang (KPZ) universality class \cite{Kardar1986,KriKru2010,Takeuchi2011,Corwin2012}. By exploiting the equivalence to models of surface growth, this scenario can be described in great detail, including in particular the non-Gaussian shape of the fitness distribution. In section \ref{ratchet}, we modify the model to study Muller's ratchet. We find that for certain parameters the rate of fitness decline does not go to zero as the population size becomes large, and we characterize the transition between fitness decline and no decline. The model with deleterious mutations is similar to a different class of models in surface growth physics, and we use this analogy to find other asymptotic properties. 

\section{Model}
\label{model}
The spatial constraints are realized as a one dimensional lattice
of size $L$ with periodic boundary conditions, where each point represents
a single organism that occupies a space \cite{Otwinowski2011a}. The evolution follows standard haploid asexual
Wright-Fisher dynamics in discrete generations, where the next
fitness of each site is chosen randomly from one of the parents in
the neighborhood, weighted according to their fitness. The smallest
possible neighborhood in one dimension is such that the child in the
next generation inherits its genotype and fitness from only two possible parents,
that is, the fitness $f_{i}(t+1)$ of site $i$ at generation $t+1$ is chosen 
from either $f_{i}(t)$ or $f_{i+1}(t)$. In other words,
$f_{i}(t+1)=f_{i}(t)$ with probability
$f_{i}(t)/(f_{i}(t)+f_{i+1}(t))$, and $f_{i}(t+1)=f_{i+1}(t)$ with
probability $f_{i+1}(t)/(f_{i}(t)+f_{i+1}(t))$.
Simulations were written in \texttt{C} and parallelized with \texttt{parallel} \cite{Tange2011a} (code available upon request).

In the case of a homogeneous system of fitness 1, where a single mutant
appears with fitness $1+s$, the fixation probability for a beneficial
mutation is the same as in the well-mixed case, $\pi=2s$ 
for $s \ll 1$ \cite{Maruyama1970,Maruyama1974}.
Intuitively, the fixation probability is unaffected because a single
mutation has ample time to compete with the entire system, regardless
of spatial structure. Since the fixation probability is the same,
the speed of evolution in the periodic selection regime is the same as in the
well-mixed case. What is different is the timescale of fixation.
The boundary between two domains with different fitnesses is a biased
random walker, and the speed of this walker is the expected value
of its displacement after one time step, $c=s/2$ for small $s$. In the
continuum limit, this model corresponds to a special case of the more
general stochastic Fisher equation (or SFKPP equation) \cite{DOERING2003,Hallatschek2009,Hallatschek2010},
where it is possible to have traveling waves with speed $c\sim s$
in the strong noise regime, or $c\sim\sqrt{s}$ in the weak noise
regime. However, the dependence of the wave speed on $s$ does not
change the essential features.
Importantly, the time for fixation may be much longer in the presence
of spatial structure compared to well-mixed populations. A wave
spreading with finite speed $c$ will take time $t_{\textrm{fix}}\sim L/c$
to cover the whole system (and total population size $N\sim L$),
as opposed to a well-mixed population where $t_{\textrm{fix}}\sim\log(N)$.
The slow spread of mutations makes it more likely that many clones exist simultaneously
in large systems. A site may also contain more than one organism,
in which case $c$ is different, but it does not change the overall
results \cite{Martens2011b} (unless interference happens within one site).

Since we are interested in the rate of evolution during competition,
a steady rate of beneficial mutations is supplied, akin to a population
adapting to a new environment. Beneficial mutations appear randomly
at rate $U_b$ per site per generation (deleterious mutations are
studied in section \ref{ratchet}). We assume that mutations have
independent effects, with no epistasis, and therefore increase the
fitness according to $\log f'=\log f+s$, where $s$ is a constant with $|s|\ll1$.

An important quantity is the rate of fitness changes $V=\lim_{t\to\infty} \langle \textrm{log} f\rangle/t$, where the average is over the population.
When the time between mutations to appear and become established,
$t_{\textrm{mut}}=(\pi U_bL)^{-1}$, is much longer than $t_{\textrm{fix}}$,
$V$ is mutation limited: $V=s\pi U_bL=2s^{2}U_bL$.
However, when $t_{\textrm{mut}}\sim t_{\textrm{fix}}$, multiple unfixed
mutations in the population compete with each other, slowing down
$V$. In well mixed populations the condition for mutation limited
adaptation is that there should be less than one new beneficial mutation
per generation. In contrast, with spatial structure $t_{\textrm{mut}}\sim t_{\textrm{fix}}$
defines a characteristic interference length scale $L_c \sim (c/U_b)^{1/2}$, 
above which mutation competition
sets in. In this competitive regime, the rate of evolution no longer
depends on the supply of beneficial mutations, but $V$ becomes independent
of $L$ for $L > L_c$ \cite{Otwinowski2011a,Martens2011b}. Using
this observation and dimensional analysis, one may deduce that this
maximum speed grows as $U_b^{1/2}$ in one dimension, and $U_b^{1/3}$ in two dimensions.

\section{Adaptation with many beneficial mutations}
\label{adaptation}

\subsection{Analogy to surface growth}

The rough spatial profile of the fitness resembles a typical surface
seen in surface growth models \cite{Otwinowski2011a,Barabasi1995}. In surface growth, 
particles are deposited on an initially smooth surface randomly,
and they may diffuse or stick to each other, gradually forming a rough
surface. Many simple models of surface growth were studied by statistical
physicists interested in non-equilibrium systems \cite{Barabasi1995,Krug1997}. They discovered
that a large number of models share the same properties in the continuum,
long-time limit, where many of the microscopic details of the model
do not matter, and these classes of models, or universality classes,
share the same symmetries.

\begin{figure}
\begin{centering}
\includegraphics[width=1\textwidth]{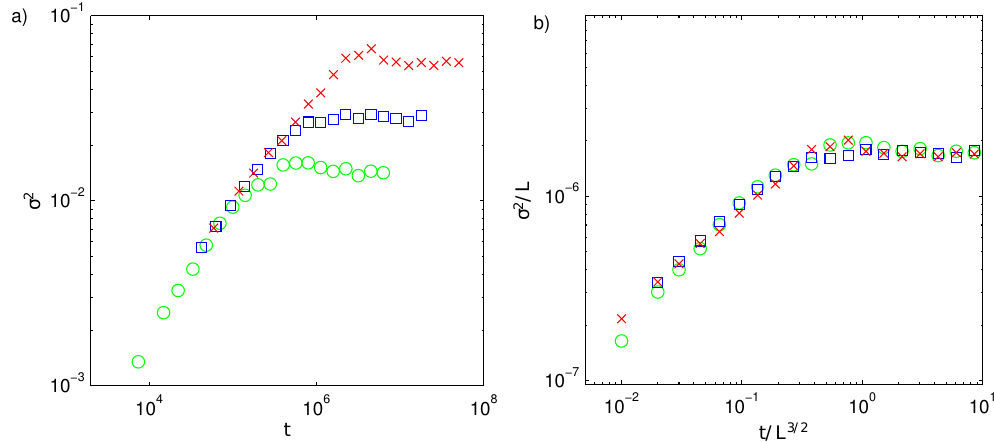}
\par\end{centering}

\caption{\label{fig:sigma}(a) Variance $\sigma^2$ of the log fitness distribution
as a function of time for different system sizes, $L=2^{13}$ (green circles),
$L=2^{14}$ (blue squares), and $L=2^{15}$ (red crosses), with $s=0.05$ and $U_b=10^{-5}$.
After a transient regime, $\sigma^2$ saturates at a value that depends
on $L$. (b) When the data is rescaled as $\sigma^{2}/L$ and $t/L^{3/2}$
it collapses onto a single curve, indicating that in fact $\sigma(t)\sim t^{1/3}$
and $\sigma(t\to\infty)\sim L^{1/2}$ as predicted by KPZ theory. Data points are averages over 50 simulations.}
\end{figure}

The evolutionary model defined here is equivalent to a surface growth
model called polynuclear growth \cite{Frank1974,Goldenfeld1984,Krug1989,Praehofer2000}
(PNG), in the limit $s\to\infty$. In PNG, the process of surface
growth may be divided into two parts, nucleation (mutation), and spreading
(selection). Nucleation occurs with low probability at any point,
at a certain rate, $U_b$, which corresponds to adding a small block
of height to the surface (log fitness). The nucleated block then grows
laterally forming a new layer. Depending on the size of the lattice, the surface grows layer by layer (corresponding to the periodic selection regime) or the surface roughens due to multiple simultaneous nucleation events
(corresponding to clonal interference) \cite{Frank1974,Goldenfeld1984}. In the rough regime the PNG model belongs to the universality class of growth
processes described on large length and time scales by the KPZ equation,
a nonlinear stochastic partial differential equation \cite{Kardar1986,Krug1989,Praehofer2000}.
While in PNG the spreading is fast and deterministic, in the evolutionary
model it is stochastic, and the new layer may even disappear \cite{Goldenfeld1984}. The
boundaries may collide with each other, and they either annihilate
or stack up creating differences in log fitness greater than $s$.
From the point of view of surface growth it is natural to hypothesize that 
the universal features of the PNG model are robust with respect to these differences, but this
has to be verified by explicit simulations. The test of the universality hypothesis 
proceeds in two steps. First, one estimates the scaling exponents governing the power law dependence of the standard deviation of the
surface height (or log fitness) distribution on time and system size. 
Second, the shape of the full distribution of height fluctuations is considered.  In surface growth, starting from flat initial conditions,
the standard deviation of the surface height distribution grows
in time as $\sigma(t)\sim t^{\beta}$, where $\beta$ is the growth
exponent, then reaches a steady state when the correlation length
reaches the size of the system \cite{Barabasi1995,Family1985}. In
the steady state, $\sigma(t\to\infty)\sim L^{\alpha}$ where $\alpha$
is the saturation exponent. Figure \ref{fig:sigma}a confirms this scenario for the evolution model.
The crossover time is where saturation sets in (the
elbow), and it scales as $L^{\alpha/\beta}$. One may try to measure
the exponents from the simulations, but based on the similarity to the PNG model 
one expects that the scaling exponents are those of the one-dimensional KPZ-equation,
$\alpha=1/2$, $\beta=1/3$ and $\alpha/\beta = 3/2$. Figure \ref{fig:sigma}b shows that
the data indeed collapses when plotted as $\sigma^{2}/L$ versus $t/L^{3/2}$. In the evolutionary 
context the saturation time scale $\sim L^{3/2}$ is proportional to the fixation time of 
beneficial mutations \cite{Martens2011b}. Note that these values of the 
exponents characterize the asymptotic, long time and large scale behavior of the model, and the behavior in the 
pre-asymptotic regime may be somewhat different \cite{Otwinowski2011a}.

\subsection{Transient fitness distributions}

Over the past decade, a much more refined characterization of the KPZ universality class has been developed
that extends beyond the values of the scaling exponents $\alpha$ and $\beta$ to the full probability distribution
of surface height fluctuations \cite{KriKru2010,Takeuchi2011,Corwin2012}. 
The essence of this refined universality hypothesis is that the log fitnesses (or surface heights) can be written as 
\begin{equation}
\log f_{i}(t)=Vt+(\Gamma t)^{1/3}\chi,\label{eq:height}
\end{equation}
where $\chi$ is a random variable from one of the \textit{Tracy-Widom (TW) distributions},
$V$ is the long-time growth rate, and $\Gamma$ is a constant related
to the parameters of the KPZ equation \cite{KriKru2010}. From eq. (\ref{eq:height}) we
find the width of the distribution:
\begin{equation}
\sigma^{2}(t)=\textrm{var}(\log f_{i}(t))=(\Gamma t)^{2/3}\textnormal{var}(\chi).\label{eq:var}
\end{equation}
The TW distributions were first discovered in fluctuations
of the largest eigenvalues of random matrices \cite{Tracy1994}.
The relation to the PNG model was established by mapping the PNG surface height 
to the length of the longest increasing subsequence of random permutations \cite{Praehofer2000,Baik1999},
and subsequently TW universality was derived directly from the KPZ equation 
\cite{Corwin2012,Sasamoto2010}.
Remarkably, the distributions were found to be geometry dependent, with the flat (monomorphic) initial
condition leading to the TW distribution characteristic of 
random matrices from the Gaussian orthogonal ensemble (GOE).

\begin{figure}
\begin{centering}
\includegraphics{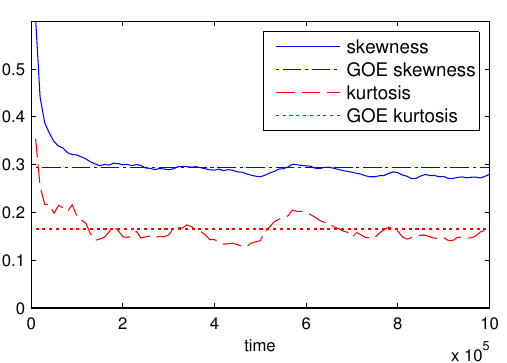}
\par\end{centering}

\caption{\label{fig:skewkurt}Skewness and kurtosis of the fitness distributions
from 200 simulations compared to the known values for the GOE Tracy-Widom distribution. $L=2^{14}$, $s=0.05$, $U_b=10^{-5}$.}
\end{figure}

Here we show numerically that, despite the additional randomness of
the stochastic spreading, the distribution of fitnesses in the non-stationary
regime of the spatial evolution model is a TW distribution characteristic
of the KPZ universality class. One signature of the TW distributions can
be seen by measuring higher moments, such as skewness, $\langle\left(\frac{\log f-\langle\log f\rangle}{\sigma}\right)^{3}\rangle$ 
and excess kurtosis, $\langle\left(\frac{\log f-\langle\log f\rangle}{\sigma}\right)^{4}\rangle-3$,
which do not depend on the parameters $V$ and $\Gamma$.
Figure \ref{fig:skewkurt} shows that the skewness and kurtosis of
the fitness distributions are non-zero, indicating non-Gaussianity,
and they approach the known values of the GOE TW distribution. 

\begin{figure}
\begin{centering}
\includegraphics[width=.6\textwidth]{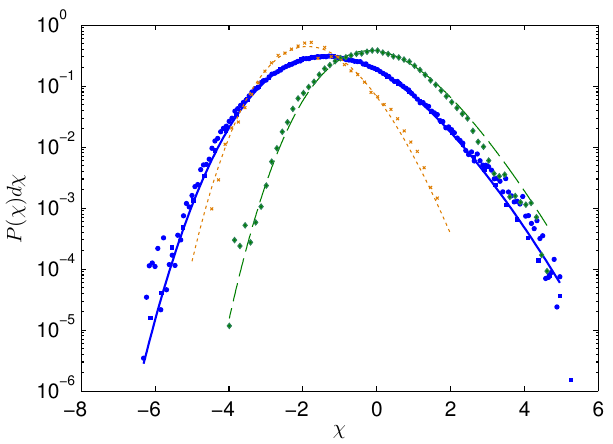}
\par\end{centering}
\caption{\label{fig:TW}
Scaled fitness distributions for three different initial
conditions: Flat (blue squares and circles), droplet (orange crosses), and rough (green diamonds). Lines indicate the Tracy-Widom GOE (blue solid), GUE (orange dotted), and the $F_{0}$ (green dashed) distributions
respectively (calculated using \cite{Dieng2006}). The scaled fitness distributions were taken from 200 simulations after $10^{6}$
generations, with $L=2^{18}$, $U_b=10^{-5}$ and $s=0.05$,
except the blue squares which had exponentially distributed selection
coefficients, with mean $\langle s\rangle=0.05$. For rough initial conditions,
the simulation was first run to the steady state ($L^{3/2}$ generations),
and deviations from the initial condition were calculated. For the droplet
geometry, a single mutation was first allowed to establish, and mutations
were only allowed in that lineage. The exact shape of the droplet
is unknown, so only fitnesses from the position of the initial mutation
(the peak of the droplet) were used in the distribution.}
\end{figure}

It is also possible to compare the fitness distribution directly to the
TW distribution. The parameters $V$ and $\Gamma$
can be found from the simulation data by applying linear regression
to the means of equations (\ref{eq:height}) and (\ref{eq:var}). The
fitnesses from the simulation are then rescaled as 
\begin{equation}
\chi_{sim}=\frac{\log f_{i}-Vt}{(\Gamma t)^{1/3}}.
\end{equation}
Figure \ref{fig:TW} shows that in the non-stationary regime, the
fitnesses fall onto the universal GOE TW distribution, which is skewed towards
higher fitnesses, with tail behaviors $ - \ln P(\chi)_{\chi\to\infty}\sim\chi^{3/2}$
and $ - \ln P(\chi)_{\chi\to-\infty}\sim \vert\chi\vert^{3}$.
To demonstrate the robustness of this result, we simulated a variant of the model where the selective
advantage of beneficial mutations, $s$, is a random variable
generated from an exponential distribution, a common choice in this field \cite{Park2010a,Good2012}.  
The two data sets can be seen to be indistinguishable.

In addition, two other initial conditions were simulated. The droplet
geometry in the PNG model is when the initial condition is a single nucleation site,
with no additional nucleations (or mutations) allowed outside. The
boundary of the initial seed grows over time, making the fitness profile
curved. The deviations from this curved profile converge to the TW
distribution of the Gaussian unitary ensemble (GUE) \cite{KriKru2010,Takeuchi2011,Praehofer2000}. The droplet geometry
has an interesting evolutionary analogy: It corresponds to a
mutation that raises the mutation rate significantly (a mutator strain),
and competes with a population that has essentially no mutations. 
The third initial condition corresponds to a system with fully developed,
stationary diversity (surface roughness). In this case the distribution of the deviations
from the initial fitness profile is predicted to converge to a universal distribution $F_{0}$,
which does not appear directly in random matrix theory but is closely related to the TW distributions 
\cite{Praehofer2000}. Again, the data fall nicely onto the predicted
distribution. The three universal distributions shown in
Fig.~\ref{fig:TW} have similar overall shapes and share the same tail behavior
mentioned above for the GOE TW distribution, but the distribution $F_0$ is distinguished from the
others by having zero mean \cite{Praehofer2000}. 

\section{Deleterious mutations}
\label{ratchet}
Our model may be modified to include only deleterious mutations by having a single negative selection coefficient $s<0$. Once a mutation appears, its expansion is unfavorable, and happens only due to genetic drift. For a single mutation, the probability of fixation vanishes exponentially in $N$ as $\pi \sim e^{Ns}$ for large negative $Ns$ \cite{Maruyama1970,Maruyama1974}. When many deleterious mutations are present 
simultaneously in the population, the rate of fitness decline is governed by the time scale on which
the subpopulation of individuals with the smallest number of mutations (the `least loaded class') goes
extinct by genetic drift. 
In the well-mixed case the number of individuals in the least loaded
class is on the order of $n_0 \sim N e^{-U_d/\vert s\vert}$, where $U_d$ is the deleterious mutation rate \cite{Haigh1978}. Correspondingly for 
\begin{equation}
\label{Muller_mixed}
n_0 \vert s \vert \sim  N \vert s \vert e^{-U_d/\vert s \vert} \gg 1.
\end{equation}
the probability of fixation of an additional deleterious mutation in this class is exponentially
small. Detailed analysis shows that under condition (\ref{Muller_mixed}) the rate of Muller's ratchet is also exponentially small in $N$ \cite{Jain2008,Waxman2010,Metzger2013}, whereas for $n_0 \vert s \vert < 1$ the fitness of the population declines continuously, and a description in terms of a traveling wave in 
fitness space, similar to that used in the context of adaptation ($s > 0$), is applicable
\cite{Rouzine2008}. Importantly, for a given set of mutation
parameters ($U_d, s$) the slow ratchet
condition (\ref{Muller_mixed}) is always attained for large populations, which implies
that the fitness decline effectively ceases for $N \to \infty$.

\subsection{Muller's ratchet in spatial populations}

\begin{figure}[htp]
\centering
\includegraphics[width=.5\textwidth]{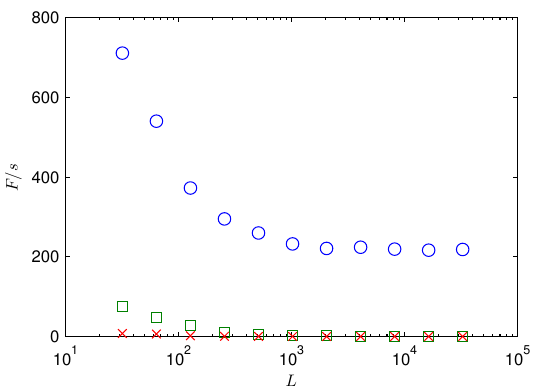}
\caption{\label{muller_N}
Deleterious mutations accumulate at a constant rate in large spatial habitats. The figure
shows the number of accumulated deleterious mutations per site, or $F/s$ after $10^6$ generations for $s=-0.01$,  $U_d=10^{-3}$ (blue circles),  $U_d=10^{-4}$ (green squares), and $U_d=10^{-5}$ (red crosses). 
For $U_d \leq 10^{-4}$ the number of mutations per site falls below unity for large $L$, while for
$U_d = 10^{-3}$ about 20\% of all $10^6$  $U_d$ mutations that occurred have been fixed.
Results were averaged over 10 instances. 
}
\end{figure}

\begin{figure}[htp]
\centering
\includegraphics[width=.45\textwidth]{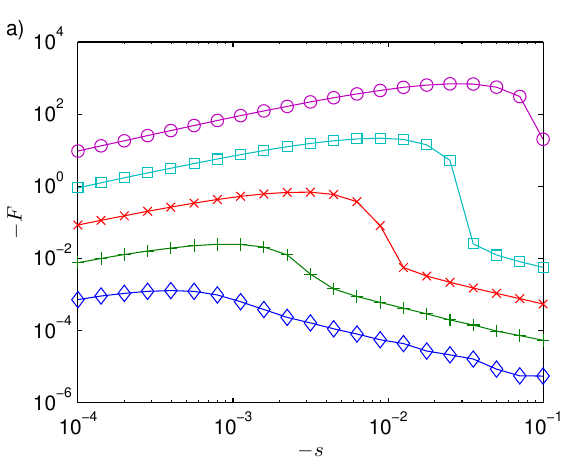}
\includegraphics[width=.45\textwidth]{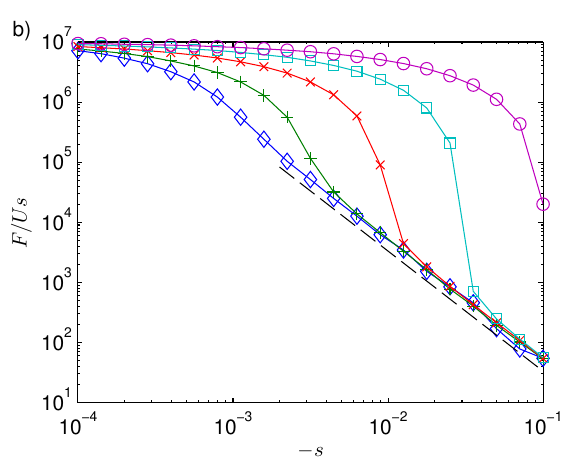}
\includegraphics[width=.45\textwidth]{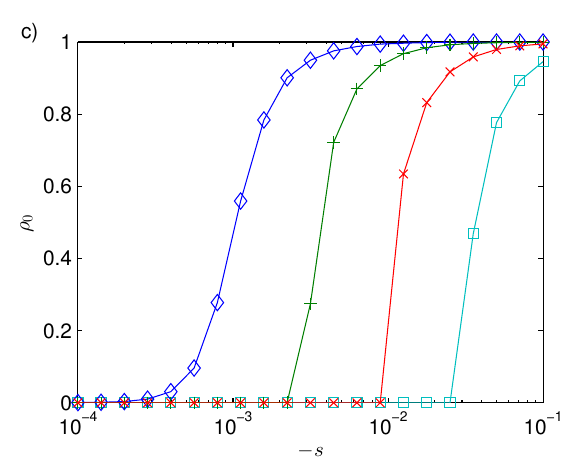}
\includegraphics[width=.45\textwidth]{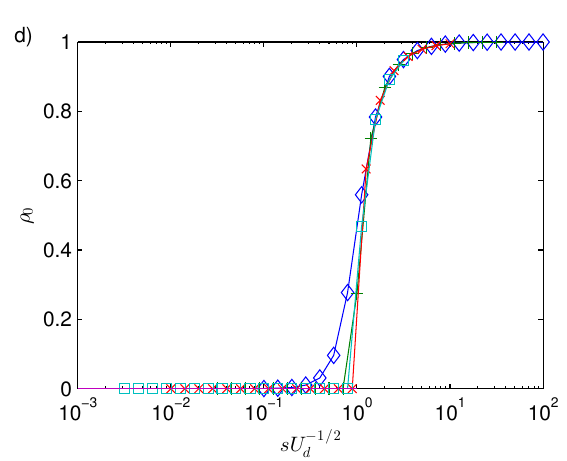}
\caption{\label{muller_US}
 (a) The rate of fitness decline, as indicated by the mean log fitness $F$, changes sharply for certain parameters $U_d$ and $s$. (b) Rescaling by $U_ds$ collapses parts of the curves. Dashed line indicates $s^{-2}$.  $F$ was measured after $10^7$ generations with $L=2^{18}$, and mutation rates were $U_d=10^{-6}$ (blue diamonds), $10^{-5}$ (green plusses), $10^{-4}$ (red crosses), $10^{-3}$ (cyan squares), and $10^{-2}$ (magenta circles). (c)
A sharp transition in the density of sites with no mutations, $\rho_0$, as a function of mutation effect size. When $\rho_0$ is large the fitness does not decline, while when $\rho_0=0$ the fitness declines indefinitely. (d) Scaling the $x$-axis by $U_d^{-1/2}$ reveals that the critical parameters are $U_d/s^2 \approx 1$. Results were averaged over 30 simulations, except for $U_d=10^{-4}$ and $U_d=10^{-3}$, which were only run once.
}
\end{figure}

Simulations of the one-dimensional spatial model show a fundamentally different behavior in the rate of fitness decline, which we measure with $F= \langle \textrm{log} f\rangle$, so that $V$ is approximately $F/t$ after some long time. Fig.~\ref{muller_N} shows that for sufficiently large $U_d$ deleterious mutations accumulate
at a constant rate, which becomes independent of the habitat size, $L$, for large $L$.  
Exploration of the parameter space reveals that this transition in the
fitness decline is sharp for certain values of $U_d$ and $s$
(Fig~\ref{muller_US}). The rate of fitness decline is non-monotonic in
$\vert s \vert$. Initially the larger mutation effects lead to a higher rate of fitness decline
with increasing $\vert s \vert$, but at the same time selection becomes more effective in 
eliminating the deleterious mutations, which eventually halts the fitness decline.
Rescaling the fitness by $U_ds$ collapses the curves in the region of large $\vert s \vert$, 
where $F \sim U_d/s$, while for very small $\vert s \vert$, mutations accumulate at close to the
maximal possible rate, $F \approx U_s s t$. 

\begin{figure}[htp]
\centering
\includegraphics[width=.5\textwidth]{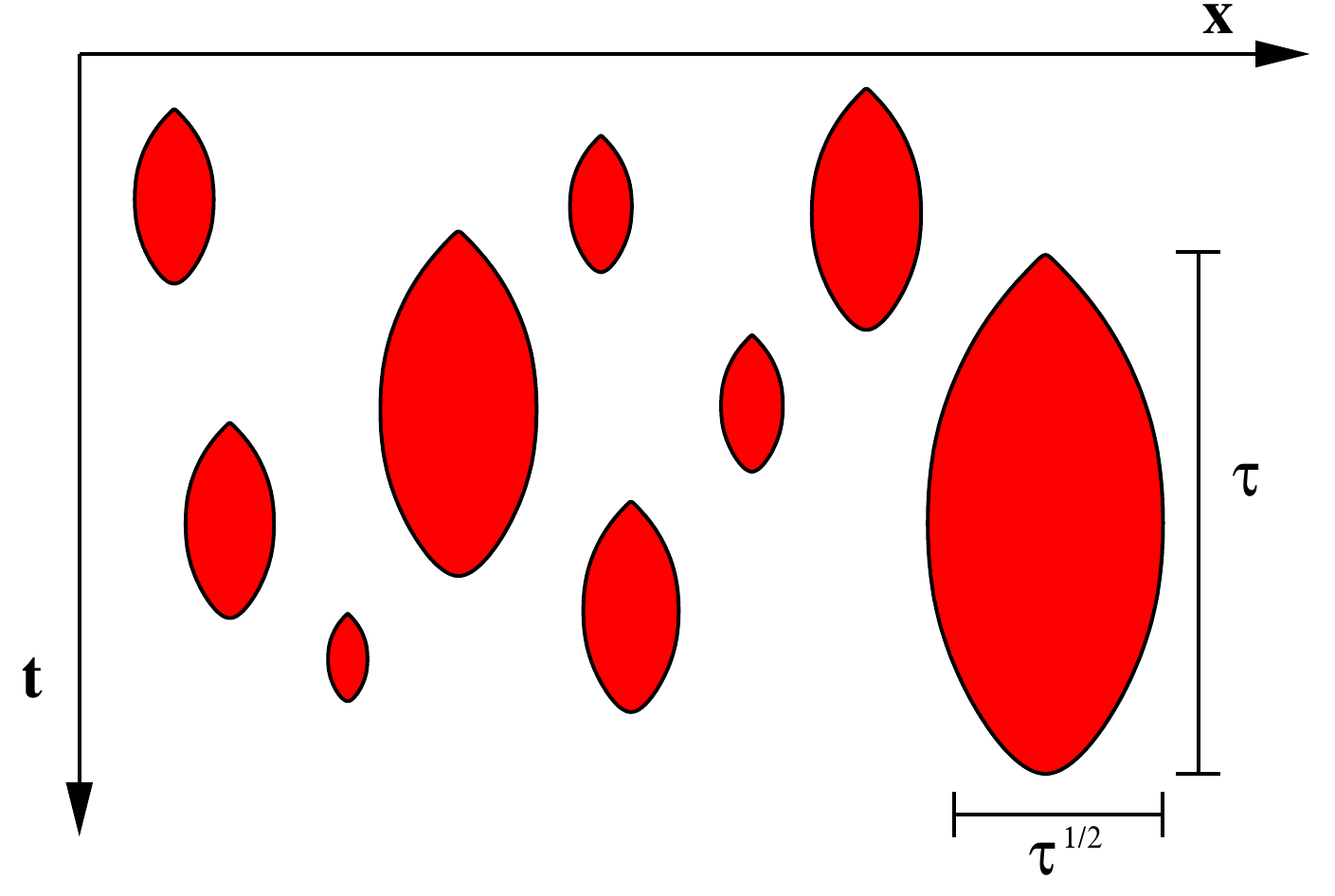}
\caption{\label{patches}
Schematic space-time view of the creation and extinction of patches of
deleterious mutations in the spatial Muller's ratchet problem. The
boundaries of a patch are weakly biased random walks and the patch
disappears when the two walks meet. A patch of life time $\tau$
reaches a maximal width $\sim \tau^{1/2}$. The distribution of life
times is heavy-tailed for small $\vert s \vert$ (see Eq.~\ref{Ptau}).   
}
\end{figure}

To further elucidate the nature of the transition we examine the
density of sites with no mutations, $\rho_0$. Figure~\ref{muller_US}c
shows a sharp transition in $\rho_0$, between 
regimes where the fitness is steadily declining (the moving ratchet)
and where the fitness is not declining (the stationary ratchet). The collapse of curves in Fig.~\ref{muller_US}d indicates that the transition occurs when 
\begin{equation}
\label{Us2}
\frac{U_d}{s^2} \approx 1.
\end{equation}
To explain this relation, consider a patch of deleterious mutants created in a single mutational 
event. Because $\vert s \vert \ll 1$, the boundaries of the patch perform almost symmetric random
walks that are weakly biased inwards by selection. The patch
disappears when the two boundaries meet. The life time $\tau$ of such
an isolated patch is therefore equal to the first passage time of a
random walk on the half-line with a bias $\sim \vert s \vert$ towards the origin, which 
has a distribution of the form \cite{Redner2001}. 
\begin{equation}
\label{Ptau}
P(\tau) \sim \tau^{-3/2} \, e^{-s^2 \tau}.
\end{equation}
When $U_d$ is small, deleterious patches are created and disappear
independently of each other (Fig.~\ref{patches}).  

\begin{figure}[htp]
\centering
\includegraphics[width=.45\textwidth]{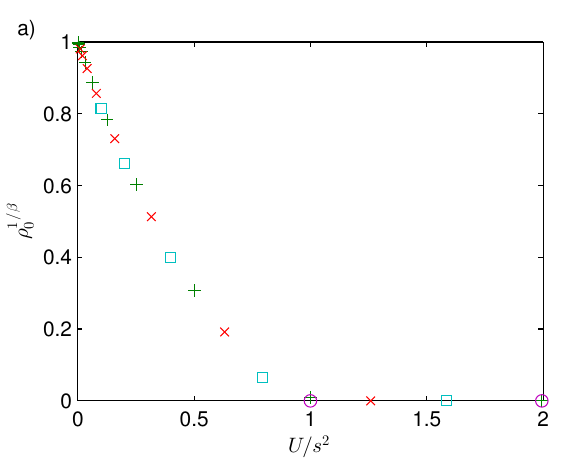}
\includegraphics[width=.45\textwidth]{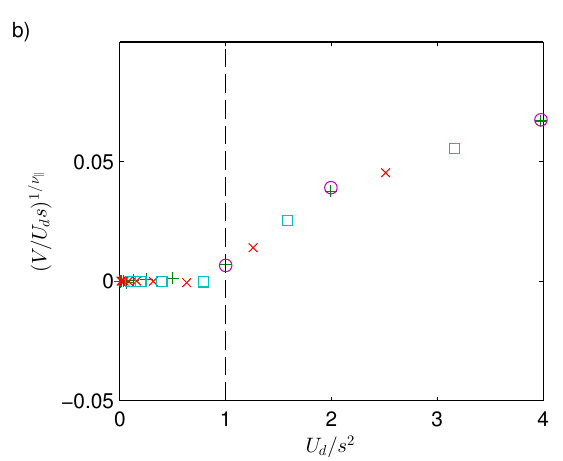}
\caption{\label{rho0_fig}
(a) The density of sites with no mutations, $\rho_0$, is a function of
$U_d/s^2$ that vanishes at $U_d/s^2 \approx 1$. The behavior of
$\rho_0^{1/\beta}$ is approximately linear at the critical point,
which is consistent with the DP prediction (\ref{DPexp}). (b)
Similarly, the scaled rate of fitness decline $V/U_ds$ in the moving phase is a function
of $U_d/s^2$ and vanishes at the transition in accordance with
(\ref{DPexp}). Dashed line indicates $U_d/s^2=1$. Simulations are described in Fig.~\ref{muller_US}. Here $V$ was estimated more accurately to exclude the genetic load, by finding the slope of $F(t)$ for the second half of simulation. Points with the lowest mutation rate, $U_d=10^{-6}$, were omitted, because they did not have enough time to reach the asymptotic velocity.
}
\end{figure}

To estimate the
density $\rho_0$ of sites without deleterious mutations, we note that
$1-\rho_0$ is the probability that a deleterious mutation is found at
a randomly chosen point in space, $x$, at a random time $t$. In other
words, $1-\rho_0$ is the fraction of the space-time area in
Fig.~\ref{patches} that is covered by deleterious patches. A patch
with life time $\tau$ reaches a spatial extension of order
$\tau^{1/2}$, and hence its area is $a \sim \tau^{3/2}$. Using
(\ref{Ptau}) the distribution of $a$ is found to be
$P(a) \sim a^{-4/3} e^{-s^2 a^{2/3}}$, from which the average
area of a patch is deduced as $\langle a \rangle \sim s^{-2}$. Since
patches are created with probability $U_d$ per unit time and space, it
follows finally that
\begin{equation}
\label{rho0}
1 - \rho_0 \sim \frac{U_d}{s^2}
\end{equation}
at least when $U_d/s^2 \ll 1$ so that the patches remain isolated. 
Assuming that the dependence of $\rho_0$ on the
parameter combination $U_d/s^2$ continues to hold up to the point
where the merging of patches leads to the global extinction of the
least loaded class ($\rho_0 = 0$), we conclude that the transition 
from the stationary to the moving ratchet is indeed 
determined by a condition of the form (\ref{Us2}). Support for this
assumption is provided in Fig.~\ref{rho0_fig}a, which shows   
that simulation results for $\rho_0$ obtained for different values of
$s$ and $U_d$ collapse onto a single curve when plotted against
$U_d/s^2$.

The relation (\ref{rho0}) also explains the behavior of the fitness in
the regime of large $\vert s \vert$ in Fig.~\ref{muller_US}. In the
stationary phase of the ratchet the fitness is independent of time and
given by $F = s (1-\rho_0) \sim U_d/s$, hence $F/ U_d s \sim 1/s^2$
independent of $U_d$. Note that the behavior of $\rho_0$ in (\ref{rho0}) 
is different from the well-mixed case, where $\rho_0 =
e^{U_d/s} \approx 1 - \frac{U_d}{\vert s \vert}$. For a given
selection strength $\vert s \vert$, the deleterious mutation rate
required to set the ratchet into motion is $U_d \sim s^2$ in the
spatial case, much smaller than the corresponding value 
$U_d \sim \vert s \vert \ln(N \vert s \vert)$ obtained from (\ref{Muller_mixed})
in the well-mixed setting. 

\subsection{Nonequilibrium wetting and critical exponents}

For a detailed characterization of the transition between the
stationary and the moving spatial ratchet we exploit the similarity of
our model to a class of surface growth models that are referred to as 
non-equilibrium wetting models \cite{Kertesz1989,Alon1998,Hinrichsen2003,Barato2009}. In
a wetting transition a macroscopic layer of one phase (typically a
liquid) forms on top of another phase (typically a solid
substrate). Non-equilibrium wetting describes the transition between a
layer that is bound to the substrate (the stationary ratchet), and one that grows indefinitely
(the moving ratchet). 
% The existence of this transition between bound and free moving
% interfaces introduces a set of associated critical exponents, in
% addition to the previous scaling exponents. 
Specifically, in the limit $s \to \infty$ our model becomes equivalent
to an unrestricted solid-on-solid model with no evaporation inside
plateaus \cite{Alon1998}. The solid-on-solid constraint implies that
the surface has no overhangs, that is, each particle is supported by another
solid particle below it. The absence of evaporation from plateaus
corresponds to the fact that fitness can increase only by selection,
and there is no restriction on the fitness/height differences between
adjacent sites. For this model the wetting transition has been shown to be
governed by directed percolation (DP) \cite{Henkel2008}.  Directed
percolation is a broad universality class of nonequilibrium phase
transitions that occur between an `active' and an `extinct' state,
for example, an infectious disease spreading in a
population. In the present context the active phase is the population
in the least loaded class that persists indefinitely in the stationary
ratchet state and goes extinct at the transition.    
% This model was shown to have unusual scaling properties, where the interface at the transition roughens logarithmically in time \cite{Hinrichsen2003}. 

\begin{figure}[htp]
\centering
\includegraphics[width=.5\textwidth]{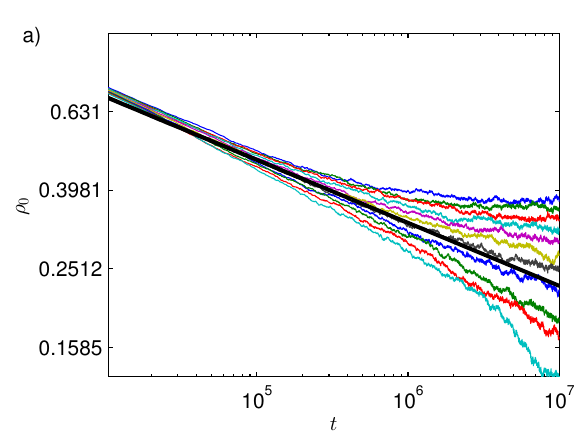}

\includegraphics[width=.45\textwidth]{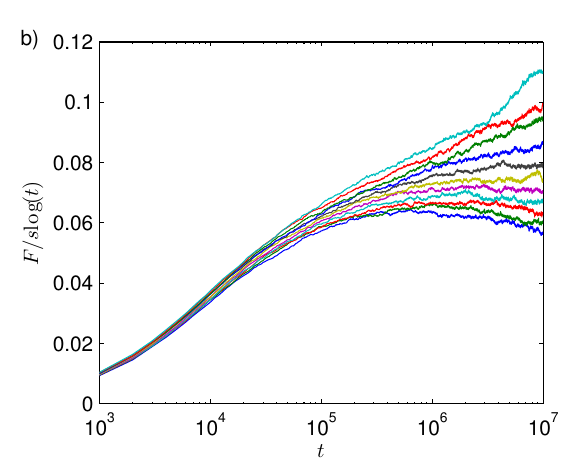}
\includegraphics[width=.45\textwidth]{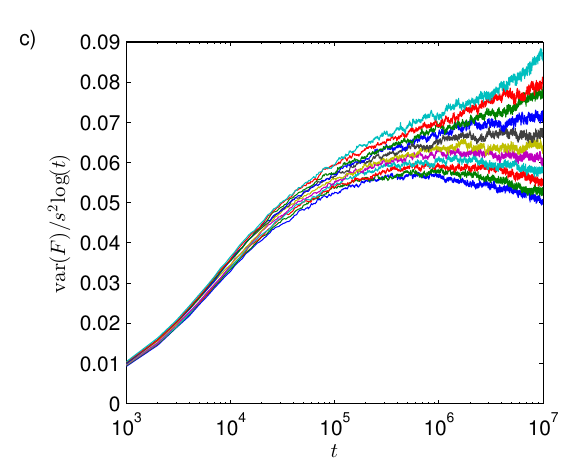}
\caption{\label{theta}
(a) Decay of the density of sites with no mutations, $\rho_0$, with time
approximately follows a power law associated with the DP class at the
critical point. High mutation rates cause $\rho_0$ to approach zero at
long times, while small amounts of deleterious mutations lead to a
non-zero value of $\rho_0$ at long times. In between, around
$U_d/s^2\approx 1$, $\rho_0(t)\sim t^{-\theta}$, with DP-exponent $\theta
\approx 0.159$ (black line). (b) Logarithmic dependence of fitness and (c) fitness variance near the critical point.
Mutation rates varied from $U_d=9\times10^{-5}$ (top line), to $U_d=1\times10^{-4}$ (bottom line), with increments of $10^{-6}$. $L=2^{18}$, $s=-0.01$, and data were averaged over 24 runs.
}
\end{figure}

The association with DP predicts power law behaviors of various
quantities near the transition.  
% There are known critical exponents associated with the DP
% transition. 
In particular, at the transition, $\rho_0(t)\sim
t^{-\theta}$, where $\theta=0.159464...$ \cite{Henkel2008}. Figure~\ref{theta} shows that $\rho_0$
decays slowly for parameters close to $U_d/s^2=1$, consistent with the power
law predicted by DP. 
Moreover, the stationary density of sites with no mutations and the rate of
fitness decline are predicted to behave as
\begin{equation}
\label{DPexp}
\rho_0 \sim (u_c -u)^{\beta_{DP}} \;\; \textrm{for} \;\; u < u_c \;\;
\textrm{and}  \;\; \vert V \vert \sim (u - u_c)^{\nu_\parallel} \;\; \textrm{for} \;\;
u > u_c, 
\end{equation}
at the transition, where $u = U_d/s^2$ is the scaled mutation rate
with critical value $u_c \approx 1$, and  $\beta_{DP} = 0.276486...$ and $\nu_\parallel =
1.733847..$ are the order parameter exponent and the temporal
correlation length exponent of DP, respectively \cite{Henkel2008}. While the additional
stochasticity associated with the smallness of the selection
coefficient in our model makes it difficult to find these exponents
numerically with any accuracy, the data shown in
Fig.~\ref{rho0_fig} are clearly consistent with the power laws (\ref{DPexp}).    
At the transition the (negative) mean fitness and the variance of the fitness are
predicted to grow logarithmically with time
\cite{Kertesz1989,Hinrichsen2003}, which is also borne out by the
simulations (Fig.~\ref{theta}b,c).

\section{Discussion}

\begin{figure}[htp]
\centering
\includegraphics[width=.5\textwidth]{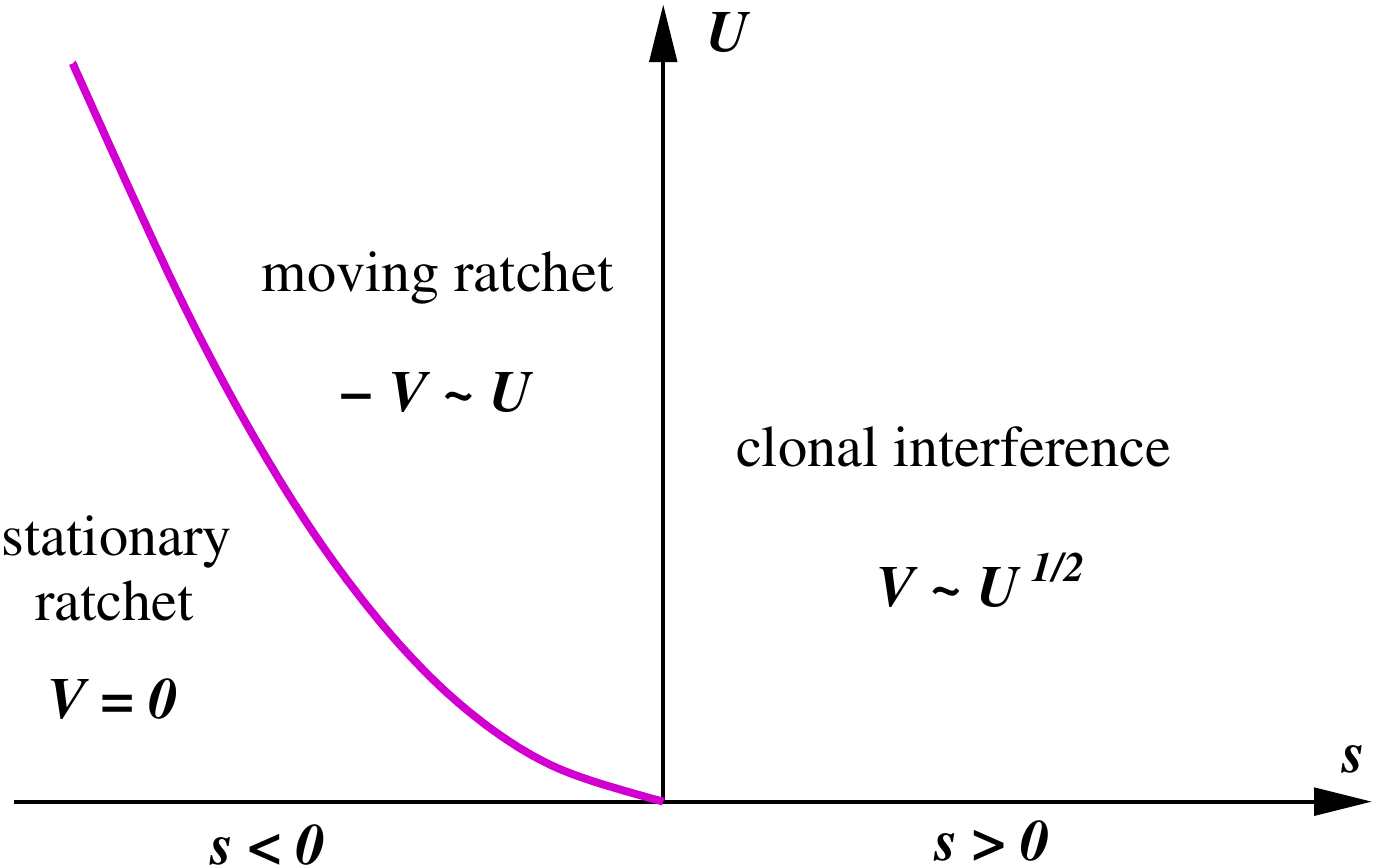}
\caption{\label{phases} Summary of evolutionary regimes as a
  function of mutation rate $U$ and selection coefficient $s$, where the latter
  is assumed to be the same for all mutations. In all regimes the rate
  of fitness change $V$ remains finite in the limit of infinite
  population size or habitat size. It is positive for adaptation ($s >
  0$), negative in the moving Muller's ratchet regime where $s < 0$
  and $U > s^2$, and zero in the stationary ratchet regime.   
}
\end{figure}

In this paper we have explored the effects of spatial structure on two
common evolutionary scenarios characterized by a large and constant supply of 
beneficial or deleterious mutations. In both
cases the fact that selection acts through local, rather than global
competition leads to profound modifications of the familiar
well-mixed dynamics. For the case of adaptation
the most conspicuous effect is the existence of a limiting
rate of adaptation that becomes independent of the population size
for large populations. At the same time the lack of communication
between different parts of the habitat implies that the fitness
variance grows without bound, invalidating the proportionality between
these two quantities expected from Fisher's fundamental theorem 
\cite{Otwinowski2011a,Martens2011b}. 
Similarly, our results for Muller's ratchet show that
selection in spatial habitats is weakened to the extent that the
fitness declines at a finite rate even for infinitely large
populations, provided the condition $U_d/s^2 > 1$ is satisfied. Figure
\ref{phases} summarizes the behavior of the rate of fitness change
in the different regimes considered in this paper. 

By exploiting analogies with models of surface growth, we have arrived
at a detailed statistical characterization of the fitness evolution in
one-dimensional spatial habitats. The model with beneficial mutations has the scaling exponents and universal distribution
that belong to the KPZ universality class, and we provide evidence
that the model with only deleterious mutations is in the directed percolation class. 
While our model becomes similar to the PNG and non-equilibrium wetting
models in the limit of strong selection, it was not a priori evident
that the additional stochasticity associated with genetic drift would leave the asymptotic behavior unchanged.

Knowing the universality class has implications for generalizations of
the model. For example, based on our understanding of KPZ-type surface
growth processes, we may conclude that the saturation of the speed of
adaptation holds in any habitat dimension and for a broad class of distributions
of selection coefficients, including those that are fatter than
exponential. Also the association between the spatial Muller's
ratchet and DP is expected to extend to two-dimensional (planar) habitats, including the
dependence of the transition on the parameter combination $U_d/s^2$.  
The scaling exponents $\beta$ and $\alpha$ governing the transient growth of the fitness
variance and its dependence on habitat size in steady state are not
known exactly for dimensions larger than one \cite{Barabasi1995,Krug1997}. However, 
a recent simulation study of various two-dimensional KPZ-models has 
identified a set of geometry-dependent universal distributions that are qualitatively similar to those found in the one-dimensional case \cite{Halpin-Healy2012}.
Spatial evolution models in planar habitats have been considered in the context of cancer progression, where 
the distribution of waiting times $t_k$ until the occurrence of a given number $k$ of mutations is of central interest \cite{Martens2011a}. 
In the surface growth analogy, this corresponds to the time when the surface reaches a given height. 
Using the probabilistic concept of first passage percolation, it can be shown that such waiting times in KPZ-type growth processes again
follow KPZ statistics \cite{KriKru2010}. This implies that the distribution of `waiting times to cancer', which was argued in \cite{Martens2011a} to be Gaussian for small $k$,
should asymptotically approach the two-dimensional analogue of the TW distribution found in \cite{Halpin-Healy2012}. 

A natural open question concerns the behavior of spatial populations
that can acquire both deleterious and beneficial mutations. In the
well-mixed case it is known that beneficial mutations dominate the
behavior of large populations, in the sense that the fitness increases
at a positive rate provided that a finite fraction of mutations are
beneficial \cite{Yu2010,Kelly2013}. In preliminary simulations    
we have explored a one-dimensional model where both types of mutations
occur at rates $U_b$ and $U_d$, respectively, with a single selection
magnitude $|s|=0.01$. 
When $U_d$ is small 
% ($U_d=1\times10^{-5}$, $U_b=1\times10^{-5}$), 
the deleterious mutations do not accumulate, but do provide a genetic load. The
genetic load does not affect the adaptation of beneficial
mutations, and the associated growth exponents and fitness
distribution are the same as for the model without any deleterious
mutations. When $U_d$ is larger, there is a competition between the
accumulation of deleterious and beneficial mutations, and the fitness
may go either up or down. Nevertheless, also in this situation the growth
exponents are close to their KPZ values, even if the fitness
is declining. A detailed investigation of this model in the light of
the analogy to non-equilibrium wetting processes appears to be an interesting
problem for further research. 

Our one dimensional model is similar to the one dimensional frontier
of an expanding planar population. Theoretical and experimental
studies have found enhanced genetic drift in such populations due to
smaller population densities at the front. This was called gene or
allele surfing \cite{Edmonds2004,Klopfstein2006,Hallatschek2007}, and
the literature is mostly concerned with the decreased genetic
variation as a signature of recent expansions. Hallatschek and Nelson
\cite{Hallatschek2010} studied the accumulation of deleterious
mutations on an expanding front and found a genetic load analogous to
eq.~(\ref{rho0}) and a sharp transition where the deleterious mutants
take over (see \cite{Peischl2013} for another study of ``expansion
load''). If separate mutations of the same phenotypic (or fitness) effect are distinguishable genetically, known as parallel adaptation, then they may also form patterns of interfering spatial waves \cite{Ralph2010}. 

An experimental test of our results would be difficult in a model
system such as expanding \textit{Escherichia coli} colonies, because
the physical growth of the colony may induce additional effects, such as super-diffusive
motion of the boundaries between genetic clones \cite{Hallatschek2007}. Also, the success of a beneficial mutation may depend on the inflation of the edge of the colony \cite{Lavrentovich2013}, or where the mutation occurs relative to the edge \cite{Lehe2012}.
Kuhr et al. \cite{Kuhr2011} modeled an expanding population with two
types of cells and unidirectional mutations. They found that the
spatial roughening of the colony boundary
changes the critical behavior of the transition to where deleterious 
mutants invade the front, compared to the directed percolation behavior
described above for the spatial Muller's ratchet.  Similarly, Lavrentovich et al.~\cite{Lavrentovich2013} found that in a radially expanding population with deleterious mutations, there was a DP-like transition whose properties are modified by the radial expansion.
% However, our model does not represent a physical moving wave of
% organisms, and our results depend only on the effects of clonal
% interference. In principle, it may apply to an  non-expanding
% microbial community adapting to new environmental conditions. 
Our model provides a theoretical understanding of the dynamics of
beneficial and deleterious mutations in one-dimensional habitats that
is separate from these additional complications.

% Finally, our model may be related to models of parallel adaptation or soft sweeps, where different mutations lead to the same phenotypic effect. In our model we only track phenotypes (fitness), and mutations arising in different locations are treated equivalently.

\subsection*{Acknowledgements} This work was supported by DFG within
SFB 680 and SPP 1590. J.O. acknowledges support from The Halle Foundation and US Army Research Office Grant W911NF-12-1-0552 awarded to Joshua B. Plotkin. We thank Oskar Hallatschek and Stefan Boettcher for useful discussions.

%\bibliographystyle{apsrev}
%\bibliography{../../library}

\end{document}